\def\year{2020}\relax
\documentclass[letterpaper]{article} 
\usepackage{aaai20}  
\usepackage{times}  
\usepackage{helvet} 
\usepackage{courier}  
\usepackage[hyphens]{url}  
\usepackage{graphicx} 
\usepackage{graphicx}  
\newcommand{\citealp}[1]{\citeauthor{#1}}
\usepackage{booktabs}       
\usepackage{amsfonts}       
\usepackage{nicefrac}       
\usepackage{microtype}      
\usepackage{xcolor}
\usepackage{multirow}
\usepackage{amsmath}
\usepackage{algorithm}
\usepackage{algpseudocode}
\usepackage{rotating}

\title{A Graph Auto-Encoder for Haplotype Assembly and Viral Quasispecies Reconstruction}
\author{
Ziqi Ke and Haris Vikalo \\
Department of Electrical and Computer Engineering \\
The University of Texas at Austin\\
ziqike@utexas.edu, hvikalo@ece.utexas.edu
}

\begin{document}

\maketitle

\begin{abstract}
Reconstructing components of a genomic mixture from data obtained by means of DNA sequencing 
is a challenging problem encountered in a variety of applications including single individual haplotyping and studies of viral communities. High-throughput DNA sequencing platforms oversample mixture components to provide massive amounts of reads whose relative positions can be determined by mapping the reads to a known reference genome; assembly of the components, 
however, requires discovery of the reads' origin -- an NP-hard problem that the existing methods struggle to solve with the required level of accuracy. In this paper, we present a learning framework based on a graph auto-encoder designed to exploit structural properties of sequencing data. The algorithm is a neural network which essentially trains to ignore sequencing errors and infers the posterior probabilities of the origin of sequencing reads. Mixture components are then reconstructed by finding consensus of the reads determined to originate from the same genomic component. Results on realistic synthetic as well as experimental data demonstrate that the proposed framework reliably assembles haplotypes and reconstructs viral communities, often significantly outperforming state-of-the-art techniques.  Source codes and datasets are publicly available at https://github.com/WuLoli/GAEseq.
\end{abstract}

\section{Introduction}

Genetic makeup of a biological sample, inferred by means of DNA sequencing, will help determine an individual's susceptibility to a broad range of chronic and acute diseases, support the discovery of new pharmaceutical 
products, and personalize and improve the delivery of health care. However, before the promises 
of personalized medicine come to fruition, efficient methods for accurate inference of genetic variations from massive DNA sequencing data must be devised. 

Information about variations in an individual genome is provided by haplotypes, ordered lists of single 
nucleotide polymorphisms (SNPs) on the individual's chromosomes (\citealp{schw2010}). High-throughput DNA sequencing 
technologies generate massive amounts of reads that sample an individual genome and thus enable studies 
of genetic variations (\citealp{schw2010,clark2004role,sabe02}). 
Haplotype reconstruction, however, remains challenging due to limited lengths of 
reads and presence of sequencing errors (\citealp{hash18}). Particularly difficult is the assembly of haplotypes in polyploids,
organisms with chromosomes organized in $k$-tuples with $k>2$, where deep coverage is typically required 
to achieve desired accuracy. This implies high cost and often renders existing haplotype assembly techniques 
practically infeasible (\citealp{motazedi2018exploit}).

A closely related problem to haplotype assembly is that of reconstructing viral communities. RNA viruses such 
as hepatitis, HIV, and Ebola, are characterized by high mutation rates which give rise to communities of 
viral genomes, the so-called viral quasispecies. Determining genetic diversity of a virus is essential 
for the understanding of its origin and mutation patterns, and the development of effective drug treatments. 
Reconstructing viral quasispecies (i.e., {\it viral haplotypes}, as we refer to them for convenience) is even 
more challenging than haplotype assembly (\citealp{ahn2018viral}) since the number of 
constituent strains in a community is typically unknown, and its spectra (i.e., strain frequencies) non-uniform.

\footnotetext[1]{This work was funded in part by the NSF grant CCF 1618427.}

Existing methods often approach haplotype assembly as the task of grouping sequencing reads according to their
chromosomal origin into as many clusters as there are chromosomes. Separation of reads 
into clusters is rendered challenging by their limited lengths and the presence of sequencing errors (\citealp{hash18}); such 
artifacts create ambiguities regarding the origin of the reads. The vast majority of existing haplotype assembly 
methods attempt to remove the aforementioned ambiguity by altering or even discarding the data, leading 
to minimum SNP removal (\citealp{lanc01}), maximum fragments cut 
(\citealp{duit10}), and minimum error correction (MEC) score (\citealp{lipp02}) optimization criteria.
Majority of haplotype assembly methods developed in recent years are focused on optimizing the MEC score, 
i.e., determining the smallest possible number of nucleotides in sequencing reads that should be altered such 
that the resulting dataset is consistent with having originated from $k$ haplotypes ($k$ denotes the ploidy of 
an organism) (\citealp{xie16,piro15,kule14,patt15,boni16}). These include the branch-and-bound scheme
(\citealp{wang05}), an integer linear programming formulation in (\citealp{chen13}), and a dynamic 
programming framework in (\citealp{kule14}). All these techniques attempt to find exact solution to the MEC 
score minimization problem; the resulting high complexity has motivated search for 
computationally efficient heuristics. They include the greedy algorithm in (\citealp{levy07}) and 
methods that compute posterior joint probability of the alleles in a haplotype sequence 
via MCMC (\citealp{bans08}) and Gibbs (\citealp{kim07}) sampling. A max-cut algorithm for haplotype 
assembly in (\citealp{bans08}) is motivated by the clustering interpretation of the problem. The
efficient algorithm proposed there, HapCUT, has recently been upgraded as HapCUT2 
(\citealp{edge17}). In (\citealp{agui12}), a novel flow-graph approach to haplotype assembly was 
proposed, demonstrating performance superior to state-of-the-art methods. More recent methods 
include a greedy max-cut approach in (\citealp{duit11}), convex optimization framework in (\citealp{das15}), 
and a communication-theoretic motivated algorithm in (\citealp{pulj16}).

Haplotype assembly for polyploids ($k>2$) is more challenging than that for 
diploids ($k=2$) due to a much larger space of possible solutions to be searched. Among 
the aforementioned methods, only HapCompass (\citealp{agui12}), SDhaP (\citealp{das15}) and BP 
(\citealp{pulj16}) are capable of solving the haplotype assembly problem for $k > 2$. Other techniques 
that can handle reconstruction of haplotypes for both diploid and polyploid genomes include a
Bayesian method HapTree (\citealp{berg14}), a dynamic programming method H-PoP (\citealp{xie16})
shown to be more accurate than the techniques in (\citealp{agui12,berg14,das15}), 
and the matrix factorization schemes in (\citealp{cai16,hash18}).

On another note, a number of viral quasispecies reconstruction methods were
proposed in recent years. Examples include ShoRAH  (\citealp{zagordi2011shorah}) and ViSpA 
(\citealp{astrovskaya2011inferring}) that perform read clustering and read-graph path search, 
respectively, to identify distinct viral components. QuasiRecomb (\citealp{topfer2013probabilistic})
casts the problem as the decoding in a hidden Markov model while QuRe (\citealp{prosperi2012qure}) 
formulates it as a combinatorial optimization. PredictHaplo (\citealp{prabhakaran2014hiv}) employs
non-parametric Bayesian techniques to automatically discover the number of viral strains in
a quasispecies. More recently, aBayesQR (\citealp{ahn2017abayesqr}) approached viral quasispecies
reconstruction with a combination of hierarchical clustering and Bayesian inference while 
(\citealp{ahn2018viral}) relies on tensor factorization.

In this paper, we propose a first ever neural network-based learning framework, named GAEseq, to both haplotype assembly and viral quasispecies reconstruction 
problems.  The framework aims to estimate the posterior probabilities of the origins of sequencing reads using an auto-encoder whose design incorporates salient characteristics of the sequencing data. Auto-encoders (\citealp{Fukushima1975}) are neural networks that in an unsupervised manner learn a low-dimensional representation of data; more specifically, they attempt to perform 
a dimensionality reduction while robustly capturing essential content of high-dimensional data 
(\citealp{goodfellow2016deep}). Auto-encoders have shown outstanding performance in a variety of applications 
across different fields including natural language processing (\citealp{richard2011}), collaborative filtering (\citealp{rianne2017graph}), and information retrieval (\citealp{thomas2016variational}), to 
name a few. Typically, auto-encoders consist of two blocks: an encoder and a decoder. The encoder converts
input data into the so-called codes while the decoder reconstructs the input from the codes. The act of copying the 
input data to the output would be of little interest without an important additional constraint -- namely, the 
constraint that the dimension of codes is smaller than the dimension of the input. This enables auto-encoders to 
extract salient features of the input data. For both the single individual and viral haplotype reconstruction problems, 
the salient features of data are the origins of sequencing reads. In our work, we propose a graph 
auto-encoder architecture with an encoder featuring a softmax function placed after the dense layer 
that follows graph convolutional layers (\citealp{Masci2011}; \citealp{rianne2017graph}); the softmax function acts as an estimator 
of the posterior probabilities of the origins of sequencing reads. The decoder assembles haplotypes by finding the 
consensus sequence for each component of the mixture, thus enabling end-to-end solution to the reconstruction 
problems.

\section{Methods}

\subsection{Problem formulation}
\label{PF}
Let $H$ denote a $k\times n$ haplotype matrix where $k$ is the number of (single individual or viral) haplotypes
and $n$ is the haplotype length. Furthermore, let $R$ denote an $m \times n$ SNP fragment matrix whose rows 
correspond to sequencing reads and columns correspond to SNP positions. Matrix $R$ is formed by first aligning 
reads to a reference genome and then identifying and retaining only the information that the reads provide about 
heterozygous genomic sites. One can interpret $R$ as being obtained by sparsely sampling an underlying ground 
truth matrix $M$, where the $i^{th}$ row of $M$ is the haplotype sampled by the $i^{th}$ read. The sampling is 
sparse because the reads are much shorter than the haplotypes; moreover, the reads may
be erroneous due to sequencing errors. Following (\citealp{ahn2018viral}) , we formalize the sampling operation as
\begin{equation}
[\mathcal{P}_\Omega (M)]_{ij} =  \left\{ \begin{array}{ll} M_{ij}, & (i,j) \in \Omega\\ 
0, & \text{otherwise}
\end{array}\right.
\end{equation}
where $\Omega$ denotes the set of informative entries in $R$, i.e., the set of $(i, j)$ such that the $j^{th}$ SNP is covered by the $i^{th}$ read, and $\mathcal{P}_\Omega$ is the projection operator denoting the sampling of haplotypes 
by reads. Sequencing is erroneous and thus $[\mathcal{P}_\Omega (R)]_{ij}$ may differ from 
$[\mathcal{P}_\Omega (M)]_{ij}$; in particular, given sequencing error rate $p$, $[\mathcal{P}_\Omega (R)]_{ij} = 
[\mathcal{P}_\Omega (M)]_{ij}$ with probability $1-p$.

Since each read samples one of the haplotypes, $R = \mathcal{P}_\Omega(UH)$ where $U$ 
denotes the $m \times k$ matrix indicating origins of the reads in $R$. In particular, each row of matrix $U$ is one 
of the $k$-dimensional standard unit vectors $e_i^{(k)}, 1\leq i\leq k$, with $1$ in the $i^{th}$ position and the 
remaining entries $0$. If $i^{th}$ read samples $j^{th}$ haplotype, the $i^{th}$ row of $U$ is $e_j^{(k)}$. If the 
origins of reads were known, each haplotype could be reconstructed by finding consensus of reads which sample 
that particular haplotype. We think of the assembly as a two-step procedure: given the SNP fragment matrix $R$ we 
first identify the read origin indicator matrix $U$ and then use $U$ to reconstruct the haplotype matrix $H$.

To characterize the performance of haplotype assembly methods we rely on two metrics: the minimum error 
correction (MEC) score, which can be traced back to (\citealp{lippert2002algorithmic}), and the correct phasing rate, 
also referred to as reconstruction rate. The MEC score is defined as the smallest number of observed entries in $R$ 
that need to be altered (i.e., corrected) such that the resulting data is consistent with having originated from $k$ 
distinct haplotypes, i.e.,
\begin{equation}
\text{MEC} = \sum_{i=1}^m \min_{j=1,2,...,k} \text{HD}(R_{i:}, H_{j:}),
\end{equation}
where HD$(\cdot,\cdot)$ denotes the Hamming distance between its arguments (sequences, evaluated only over 
informative entries), $R_{i:}$ denotes the $i^{th}$ row of $R$ and $H_{j:}$ denotes the $j^{th}$ row of $H$. 
The correct phasing rate (CPR) is defined as 
\begin{equation}
\text{CPR} = 1 - \frac{1}{kn}(\min \sum_{i=1}^k \text{HD}(H_{i:}, \mathcal M(H_{i:}))),
\end{equation}
where $\mathcal M$ is the one-to-one mapping from the set of reconstructed haplotype to the set of true haplotype (\citealp{hash18}), 
i.e., mapping that determines the best possible match between the two sets of haplotypes. To characterize performance 
of methods for reconstruction of viral quasispecies with generally a priori unknown number of components, in addition 
to correct phasing rate we also quantify {\it recall rate}, defined as the fraction of perfectly reconstructed components in 
a population (i.e., recall rate = $\frac{TP}{TP+FN}$), and {\it predicted proportion}, defined as the ratio of the estimated 
and the true number of components in a genomic mixture (\citealp{ahn2018viral}). 

To assemble haplotypes from a set of reads we design and employ a graph auto-encoder. Fig. 1 (b) shows the entire 
end-to-end pipeline that takes the collection of erroneous reads and generates reconstructed haplotypes. First, the SNP 
fragment matrix $R$ is processed by the graph encoder to infer the read origin indicator matrix $U$; then, a haplotype
decoder reconstructs matrix $H$. The graph auto-encoder is formalized in the next section.

\begin{figure*}[h!]
	\centering
		\includegraphics[width=0.9\textwidth]{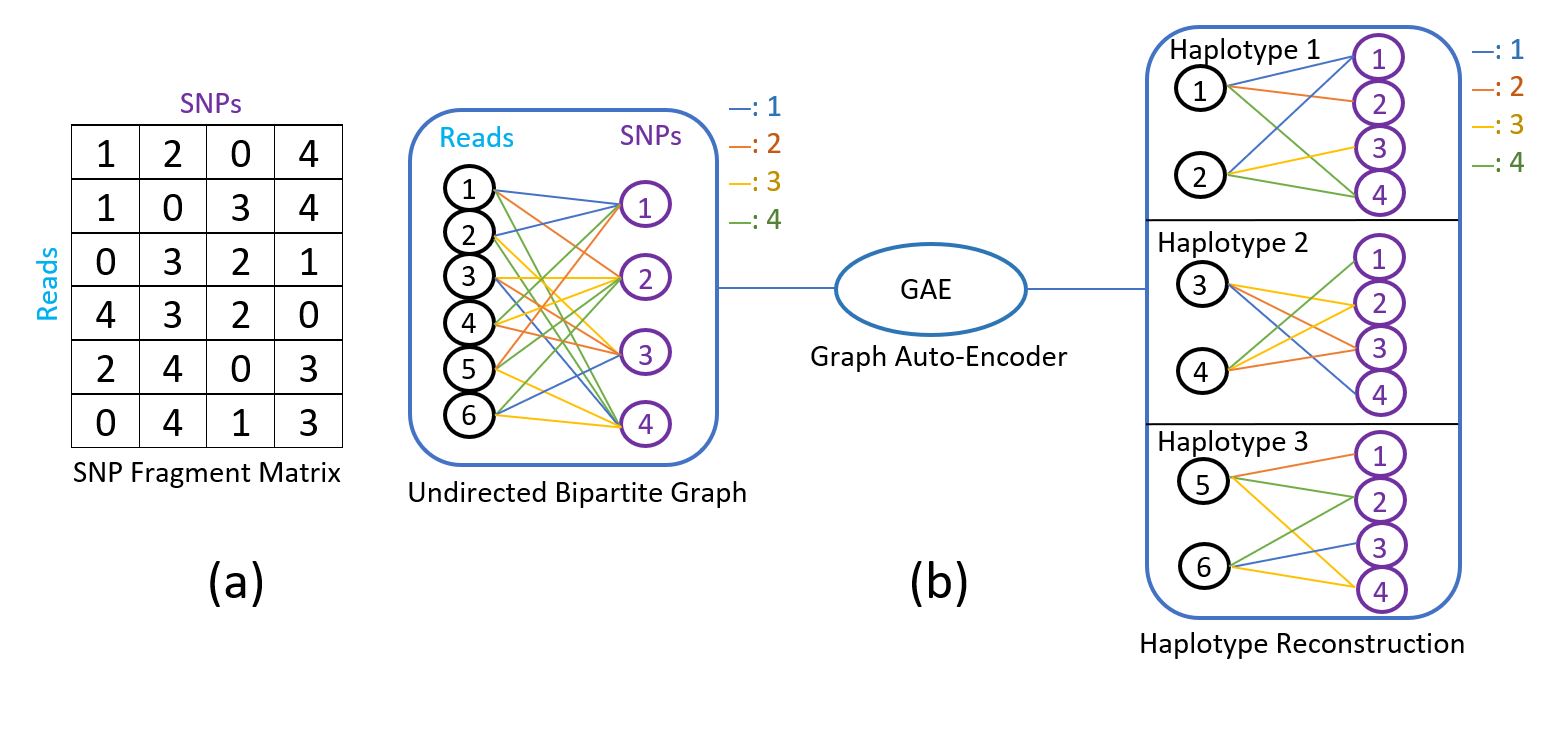}\label{fig1}
		\caption{(a) Segment of the SNP fragment matrix. Non-zero entries represent SNP information provided by sequencing reads; labels $1$-$4$ indicate the four nucleotides. Zero entries in a row indicate that the read does not cover corresponding SNP. In this illustration, the first two rows represent reads originating from the same haplotype; the third and fourth reads both originated from another haplotype; and so on. (b) The pipeline from the SNP fragment matrix to haplotypes via a graph auto-encoder.}
	\end{figure*}

\subsection{Graph auto-encoders}
Graph auto-encoders are a family of auto-encoders specifically designed for learning on graph-structured data (\citealp{rianne2017graph}; \citealp{thomas2016variational}). In this paper, we design graph auto-encoders for the assembly of the components of a genomic mixture. As in conventional auto-encoder structures, the developed architecture consists of two parts: the graph encoder and the decoder. The graph encoder $Z = f(R, A)$ takes the SNP fragment matrix $R$ and the $m \times n$ graph adjacency matrix $A$ as inputs, and outputs the $m \times k$ node embedding matrix $Z$. Note that we impose constraints on the node embedding matrix so that the salient features extracted by a graph auto-encoder approximate the read origin indicator matrix $U$. Such a constraint does not prevent efficient training of the auto-encoders via backpropagation. The decoder $\hat{R} = g(Z)$ is utilized to reconstruct the SNP fragment matrix $R$  and the haplotype matrix $H$ from the node embedding matrix $Z$; this 
implies that the decoder is essentially capable of imputing the unobserved entries in the SNP fragment matrix.

To numerically represent information in the SNP fragment matrix $R$, we encode its entries $R_{ij}$ using a set of $4$ 
discrete values -- one for each of $4$ possible nucleotides -- where the mapping between nucleotides and the discrete 
values can be decided arbitrarily. To this end, we may simply represent the nucleotides A, C, G and T by $1$, $2$, $3$ 
and $4$, respectively; non-informative entries in each row of $R$, i.e., SNP positions not covered by a read, are represented by $0$. Note that the SNP fragment matrix can be represented by an undirected 
bipartite graph $G = (V, E, \mathcal{W})$ where the set of read nodes $r_i \in \mathcal{A}$ with $i \in \{1,...,m\}$ and 
the set of SNP nodes $s_j \in \mathcal{B}$ with $j \in \{1,...,n\}$ together form the set of vertices $V$, i.e., 
$\mathcal{A} \cup \mathcal{B} = V$. The weights $w \in \{1, 2, 3, 4\} = \mathcal{W}$ assigned to edges $(r_i, w, s_j) \in E$ 
are the discrete values used to represent nucleotides. With this model in place, we can rephrase the graph encoder as 
$Z = f(R, A_1, A_2, A_3, A_4)$, where $A_w \in \{0, 1\}^{m \times n}$ represents the graph adjacency matrix for a 
nucleotide encoded by $w$. Equivalently, $A_w$ has 1's for the entries whose corresponding positions in $R$ are encoded 
by $w$. Since we are interested in imputing the unobserved entries based on the observed entries in $R$ instead of simply 
copying the observed entries to $\hat{R}$, it is beneficial to reformulate the decoder as $\hat{R} = g(Z, R)$. In other words, 
the auto-encoder is trained to learn from the observed entries in order to determine origin of reads, impute unobserved 
entries of $R$, and reconstruct haplotypes in the genomic mixture. 

\subsection{Read origin detection via graph encoder}

Recall the interpretation that the SNP fragment matrix $R$ is obtained by erroneously sampling an underlying ground truth matrix $M$. This motivates development of a specific graph encoder architecture, motivated by the ideas of the design in (\citealp{rianne2017graph}), that is capable of detecting origin of sequencing reads in $R$ via estimating the posterior probabilities of the origin of each read.

Let $D_{r}$ denote an $m \times m$ diagonal read degree matrix whose entries indicate the number of SNPs covered by each read, and let $D_{s}$ denote an $n \times n$ diagonal SNP degree matrix whose entries indicate the number of reads covering each SNP. We facilitate exchange of messages between read nodes and SNP nodes in the graph, initiating it from the set of read nodes $\mathcal{A}$; doing so helps reduce the dimensions of weights and biases since the number of reads $m$ is far greater than the haplotype length $n$. Note that the dimension of messages keeps reducing during the message passing procedure. 

\begin{figure*}[h!]
	\label{fig:02}
	\centering
		\includegraphics[width=0.9 \textwidth]{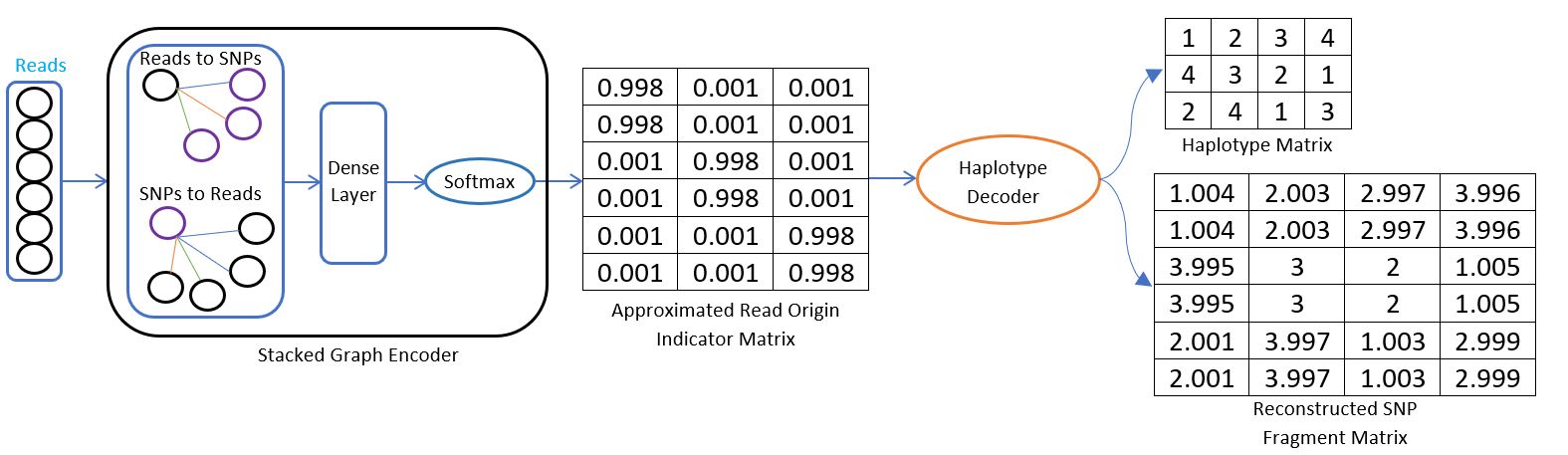}
		\caption{A forward pass through the graph auto-encoder consisting of a stacked graph encoder that passes messages between read and SNP nodes and constructs approximate read origin indicator matrix via the softmax function. Decoder reconstructs haplotypes and SNP fragment matrix.}
	\label{fig2}
\end{figure*}

The messages from read nodes to SNP nodes are 
\begin{equation} \label{r2s1}
M_{(1)} = \sigma(\sum_{w=1}^4 D_s^{-1}A_w^TRW_w^{(1)} + B_w^{(1)}),
\end{equation}
where $W_w^{(1)}$ and $B_w^{(1)}$ denote the weights and biases of the first convolutional layer for the nucleotide encoded 
with $w$, respectively, $\sigma$ denotes an element-wise activation function such as $\text{ReLU}(\cdot) = \max (\cdot, 0)$, 
and $(\cdot)^T$ denotes the transpose of a matrix. The dimension of both $W_w^{(1)}$ and $B_w^{(1)}$ is $n \times c_{(1)}$, 
where $c_{(1)}$ denotes the message length after the first message passing step. 

The messages from SNP nodes to read nodes are
\begin{equation} \label{s2r1}
M_{(2)} = \sigma(\sum_{w=1}^4 D_r^{-1}A_wM_{(1)}W_w^{(2)} + B_w^{(2)}),
\end{equation}
where $W_w^{(2)}$ and $B_w^{(2)}$ denote the weights and biases of the second convolutional layer for the nucleotide encoded 
with $w$, respectively. The dimension of both $W_w^{(2)}$ and $B_w^{(2)}$ is $c_{(1)}\times c_{(2)}$, where $c_{(2)}$ denotes 
the message length after the second message passing step.

Repeating message passing and stacking the convolutional layers leads to formation of a deep model. 
The read nodes to SNP nodes layer is readily generalized as
\begin{equation} \label{r2s}
M_{(2i+1)} = \sigma(\sum_{w=1}^4 D_s^{-1}A_w^TM_{(2i)}W_w^{(2i+1)} + B_w^{(2i+1)}),
\end{equation}
where $i \in \{0, 1, 2, ...\}$ and $M_{(2i)} = R$ for $i=0$. The dimension of $M_{(2i)}$ is $m \times c_{(2i)}$. Furthermore, the SNP 
nodes to read nodes layer is generalized as
\begin{equation} \label{s2r}
M_{(2i)} = \sigma(\sum_{w=1}^4 D_r^{-1}A_wM_{(2i-1)}W_w^{(2i)} + B_w^{(2i)}),
\end{equation}
where $i \ge 1$. The dimension of $M_{(2i-1)}$ is $n \times c_{(2i-1)}$. Note that the messages are passed from read 
nodes to SNP nodes when the subscript of $M$ is odd, and otherwise traverse in the opposite direction.

Equation (\ref{r2s}) and (\ref{s2r}) specify the graph convolutional layer while the dense layer is defined as
\begin{equation}
O = \sigma (M_{(l)}W_d + B_d),
\end{equation}
where $O$ denotes the output of the dense layer, $W_d$ and $B_d$ are the weights and biases of the dense layer, respectively, $M_{(l)}$ is the output of the last graph convolutional layer, and $l$ represents the number of graph convolutional layers. The dimension of $W_d$ is $c_{(l)} \times k$ and the dimension of $O$ and $B_d$ is $m \times k$, where $k$ denotes the ploidy
(i.e., the number of components in a genomic mixture).

To find $Z$ which approximates the read origin indicator matrix $U$ (i.e., $Z$ with each row close in the $l_2$-norm sense to a 
$k$-dimensional standard basis vector), we employ the softmax function 
\begin{equation}
Z_{ij} = \frac{e^{\beta O_{ij}}}{\sum_{j=1}^k e^{\beta O_{ij}}},
\end{equation}
where in our experiments we set $\beta$ to 200. Having estimated read origins by the node embedding matrix $Z$, the reads 
can be organized into $k$ clusters. This enables straightforward reconstruction of haplotypes by determining the consensus 
sequence for each cluster.

\subsection{Haplotype decoder}

Thus far, we have conveniently been representing alleles as the numbers in $\{1,2,3,4\}$. It is desirable, however, that 
in the definition of a loss function the distance between numerical values representing any two alleles is identical, no 
matter which pair of alleles is considered; this ensures the loss function relates to the MEC score -- the metric of
interest in haplotype assembly problems. Following (\citealp{ahn2018viral}), we define the loss function of the auto-encoder as the squared Frobenius norm of the difference between a one-hot SNP fragment matrix $\mathcal{R}$ and the reconstructed 
matrix $\hat{\mathcal{R}} = Z\mathcal{H}$ at the informative positions, i.e., 
$\mathcal{L} = \frac{1}{2} ||\mathcal{P}_\Omega(\mathcal{R} - Z\mathcal{H})||_F^2$, where 
$\mathcal{R} \in \{0, 1\}^{m \times 4n}$ and $\mathcal{H} \in \{0, 1\}^{k \times 4n}$ are formed by substituting discrete 
values $w \in \{1, 2, 3, 4\}$ by the set of four dimensional standard basis vectors $e_i^{(4)}, 1 \leq i \leq 4$. 
With such a notational convention, the proposed loss 
function approximates the MEC score; it only approximates the score, rather than coincides with it, because $Z$ is an 
approximation of the read-origin matrix $U$. Therefore, the graph auto-encoder is trained to approximately minimize 
the MEC score. Fig.~2 illustrates the data processing pipeline that takes as inputs reads in the SNP fragment matrix and 
produces the matrix of haplotypes as well as imputes missing entries in the SNP fragment matrix. The proposed graph 
auto-encoders for haplotype assembly and viral quasispecies reconstruction are formalized as Algorithm \ref{algorithm1} 
and Algorithm \ref{algorithm2}, respectively. For the viral quasispecies reconstruction problem, the number of clusters 
$k$ is typically unknown; detailed strategy based on (\citealp{ahn2018viral}) for the automated inference of $k$ can be found in Supplementary Document B.

\begin{algorithm}[h!]
	\centering
	\caption{Graph auto-encoder for haplotype assembly}\label{algorithm1}
	\begin{algorithmic}[1]
		\State \textbf{Input:} SNP fragment matrix $R$, the number of experiments $n_{exp}$ and the number of haplotpyes $k$
		\State \textbf{Output:} Reconstructed haplotypes $H$
		\While{$n_{exp} \neq 0$} 
		\State {\small Initialize $W_w^{(i)}$, $B_w^{(i)}$, $W_d$ and $B_d$ using Xavier initialization where $w \in \{1, 2, 3, 4\}$ and $i \in \{1, 2\}$}
		\For{$n_{epoch} = 1$ $\text{to}$ $100$}
		\State {\tiny $M_{(1)} \gets \sigma(\sum_w D_s^{-1}A_w^TRW_w^{(1)} + B_w^{(1)})$
			\State $M_{(2)} \gets \sigma(\sum_w D_r^{-1}A_wM_{(1)}W_w^{(2)} + B_w^{(2)})$}
		\State {\small $O \gets \sigma (M_{(2)}W_d + B_d)$
			\State $Z_{ij} \gets\frac{e^{\beta O_{ij}}}{\sum_{j=1}^k e^{\beta O_{ij}}}$ with $\beta = 200$
			\State Calculate $\mathcal{H}$ by majority voting
			\State $\mathcal{L} \gets \frac{1}{2} ||\mathcal{P}_\Omega(\mathcal{R} - Z\mathcal{H})||_F^2$
			\State Record reconstructed haplotypes and the MEC score
			\State Update $W_w^{(i)}$, $B_w^{(i)}$, $W_d$ and $B_d$ using Adam Optimizer where $w \in \{1, 2, 3, 4\}$ and $i \in \{1, 2\}$}
		\EndFor
		\State {\small $n_{exp} \gets n_{exp} - 1$}
		\EndWhile
		\State Output the reconstructed haplotypes $H$ corresponding to the lowest MEC score
	\end{algorithmic}
\end{algorithm}

\begin{algorithm}[h!]
	\centering
	\caption{Graph auto-encoder for viral quasispecies reconstruction}\label{algorithm2}
	\begin{algorithmic}[1]
		\State \textbf{Input:} SNP fragment matrix $R$, the number of experiments $n_{exp}$, the MEC improvement rate threshold $\eta$ and the estimated initial number of components $k_0$
		\State \textbf{Output:} Reconstructed viral haplotypes $H$ and the inferred frequencies
		\State Initial $\tau \gets 0$, MECflag $\gets 0$ and $k_\tau \gets k_0$
		\While{$\tau = 0$ or $k_\tau = k_\tau - 1$}
		\For{$k \in \{k_\tau, k_\tau + 1 \}$}
		\State {\small Run Algorithm $\ref{algorithm1}$ with $k$}
		\EndFor
		\If{MECimpr$(k_\tau) \le \eta $}
		\State {\small $k_{\tau+1} \leftarrow 	\lfloor(k_{\tau} + \max \{1, k_i\})/ 2 \rfloor, \{ i\in \{1,\cdots,\tau \text{-}1\} : k_i \le k_\tau \}$};
		{\small MECflag $\gets 1$}
		\Else
		\If{MECflag $= 0$}
		\State {\small $k_{\tau + 1} \gets 2k_\tau$}
		\Else
		\State {\small $k_{\tau+1} \leftarrow \lfloor(k_{\tau} + \min k_i )/2\rfloor, \{ i \in \{1,\cdots,\tau \text{-}1\} : k_i > k_\tau \}$}
		\EndIf
		\EndIf
		\State {\small $\tau \gets \tau + 1$}
		\EndWhile
		\State Output the viral quasispecies $H$ with $k = k_\tau + 1$ and the inferred frequencies
	\end{algorithmic}
\end{algorithm}  

\section{Results}

The hyper-parameters of GAEseq are determined by training on 5 synthetic triploid datasets with coverage 30$\times$ and 
validated on different 5 synthetic triploid datasets with the same coverage. The results reported in this section are obtained on test data.
Detailed description of the computational platform and the choice of hyper-parameters can be found in Supplementary Document A. 

\begin{table*}[h!]
	\caption{Performance comparison on biallelic Solanum Tuberosum 
		semi-experimental data.}
	\centering\label{tab:01}
	{\begin{tabular}{ p{0.13\textwidth} p{0.13\textwidth} p{0.13\textwidth} p{0.13\textwidth} p{0.13\textwidth} p{0.13\textwidth}}\toprule 
			& & \multicolumn{2}{l}{MEC} & \multicolumn{2}{l}{CPR}\\
			Coverage & & Mean & SD & Mean & SD\\
			\hline 
			\multirow{4}{*}{15}& GAEseq & \textbf{8.200} & \textbf{4.686} & \textbf{0.822} & 0.048\\
			& HapCompass & 100.700 & 66.150 & 0.763 & \textbf{0.046}\\
			& H-PoP & 28.700 & 32.667 & 0.783 & 0.066\\
			& AltHap & 59.100 & 28.125 & 0.709 & 0.054\\
			\hline
			\multirow{4}{*}{25}& GAEseq & \textbf{8.400} & \textbf{4.719} & \textbf{0.831} & 0.081\\
			& HapCompass & 124.800 & 132.156 & 0.810 & 0.063\\
			& H-PoP & 33.800 & 47.434 & 0.798 & \textbf{0.046}\\
			& AltHap & 92.600 & 83.649 & 0.756 & 0.068\\
			\hline
			\multirow{4}{*}{35}& GAEseq & \textbf{10.700} & \textbf{3.234} & \textbf{0.857} & 0.087\\
			& HapCompass & 217.400 & 174.135 & 0.775 & \textbf{0.072}\\
			& H-PoP & 41.700 & 53.971 & 0.823 & 0.094\\
			& AltHap & 164.000 & 101.583 & 0.754 & 0.093\\
			\bottomrule
	\end{tabular}}
\end{table*}

\begin{figure*}[h!]
	\label{fig:03}
	\begin{center}
		\includegraphics[width=0.3 \textwidth]{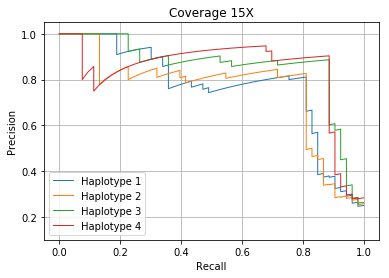}
		\includegraphics[width=0.3 \textwidth]{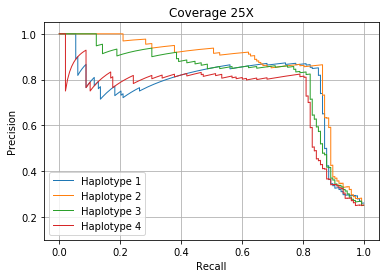}
		\includegraphics[width=0.3 \textwidth]{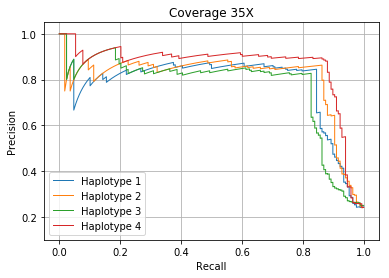}
		\caption{The precision-recall curves for Solanum Tuberosum semi-experimental data with coverage 15$\times$, 25$\times$ and 35$\times$}
	\end{center}
\end{figure*}

\begin{table*}[h!]
	\caption{Performance comparison of GAEseq, PredictHap, TenSQR and aBayesQR on a real HIV-1 5-virus-mix data. 
		Genes where all the strains are perfectly reconstructed are denoted as boldface.}
	\centering
	\label{tab:06}
	\resizebox{1 \textwidth}{!}{
		{\begin{tabular}{c c c c c c c c c c c c c c c }\toprule 
				& &  p17 &  p24&  p2-p6 &  PR&  RT &  RNase & int &  vif &  vpr &  vpu &  gp120 &  gp41 &  nef\\
				\hline
				GAEseq &  PredProp & \textbf{1} & 1 & \textbf{1} & \textbf{1} & \textbf{1.2} & \textbf{1} & \textbf{1} & \textbf{1} & \textbf{1} & 1.2 & 1 & 1 & 1\\
				&  CPR$_{ HXB2}$ &  \textbf{100} & 99.4 & \textbf{100} & \textbf{100} & \textbf{100} & \textbf{100} & \textbf{100}  & \textbf{100}  & \textbf{100} & 100 & 96.2 & 96.7  & 100 \\
				&  CPR$_{ 89.6}$& \textbf{100} & 99.4 & \textbf{100} & \textbf{100} & \textbf{100} & \textbf{100} & \textbf{100} & \textbf{100}  & \textbf{100} & 99.2 & 99.4 & 100  & 98.2\\
				&  CPR$_{ JR-SCF}$ & \textbf{100} & 100 & \textbf{100} & \textbf{100} & \textbf{100} & \textbf{100} & \textbf{100}  & \textbf{100}  & \textbf{100} & 100 & 99.9 & 100  & 99.3\\
				&  CPR$_{ NL4-3}$ & \textbf{100} & 100 & \textbf{100} & \textbf{100} & \textbf{100} & \textbf{100} & \textbf{100}  & \textbf{100}  & \textbf{100} & 100 & 100 & 100  & 99.8\\
				&  CPR$_{ YU2}$ & \textbf{100} &  100 & \textbf{100} & \textbf{100} & \textbf{100} & \textbf{100} & \textbf{100}  & \textbf{100}  & \textbf{100} & 100 & 99.6 & 100  & 98.1 \\
				PredictHap &  PredProp &  \textbf{1} &  0.6 &  \textbf{1} &  \textbf{1} &  \textbf{1} &  0.8 &  0.8 & 0.8 & \textbf{1} & 0.8 & 0.8 & 0.8 & 0.8\\
				&  CPR$_{ HXB2}$ & \textbf{100}  & 0  & \textbf{100} & \textbf{100} & \textbf{100} & 98.9  & 100 & 100 & \textbf{100} & 93.2 & 0 & 0 & 0 \\
				&  CPR$_{ 89.6}$ & \textbf{100} & 100 & \textbf{100} & \textbf{100} & \textbf{100} & 100 & 99.8 & 100 & \textbf{100} & 0 & 97.8 & 100 & 98.8\\
				&  CPR$_{ JR-SCF}$ & \textbf{100} &  100 & \textbf{100} & \textbf{100} & \textbf{100} & 100 & 100 & 100 & \textbf{100} & 100 & 99.7 & 100 & 100\\
				&  CPR$_{ NL4-3}$ & \textbf{100} & 99.1 & \textbf{100} & \textbf{100} & \textbf{100} & 100 & 100 & 100 & \textbf{100} & 100 & 100 & 100 & 100\\
				&  CPR$_{ YU2}$ & \textbf{100} & 0 & \textbf{100} & \textbf{100} & \textbf{100} & 0 & 0 & 0 & \textbf{100} & 100 & 98.6 & 100 & 100\\
				TenSQR &  PredProp & \textbf{1} & 1.6 & \textbf{1} & 1 & 1.4 &  \textbf{1}  &  \textbf{1}  &  \textbf{1}   &  \textbf{1}  & 1.6 & 2.2 & 1.2 & 0.8\\
				&  CPR$_{ HXB2}$ & \textbf{100} & 98.9 & \textbf{100} & 100 & 99.2 & \textbf{100}  & \textbf{100}  & \textbf{100}  & \textbf{100}  & 92.8 & 96.0 & 99.0 & 0\\
				&  CPR$_{ 89.6}$ & \textbf{100} & 100 & \textbf{100} & 100  & 98.0 & \textbf{100}  & \textbf{100}  & \textbf{100}  & \textbf{100}  & 94.0 & 97.2 & 100 & 95.7\\
				&  CPR$_{ JR-SCF}$ & \textbf{100} & 100 & \textbf{100} & 100 & 100 & \textbf{100} & \textbf{100} & \textbf{100} & \textbf{100} & 100 & 98.3 & 97.7 & 99.8\\
				&  CPR$_{ NL4-3}$ &  \textbf{100}& 99.3 & \textbf{100} & 100 & 99.5 & \textbf{100} & \textbf{100} & \textbf{100} & \textbf{100} & 100 & 99.8 & 99.5 & 99.7  \\
				&  CPR$_{ YU2}$ & \textbf{100} &  99.3 & \textbf{100} & 99.7 & 99.7 & \textbf{100} & \textbf{100} & \textbf{100} & \textbf{100} & 100 & 94.9 & 100 & 98.6\\
				aBayesQR & PredProp & \text{1} & 1 & \textbf{1}  & \textbf{1}  & 1 & \textbf{1}  & 1 & 1 & \textbf{1.2} & 1 & 0.8 & 0.8 & 1.2\\
				&  CPR$_{ HXB2}$ & \textbf{100} & 99.4 & \textbf{100} & \textbf{100} & 98.5 & \textbf{100} & 99.9 & 100 & \textbf{100} & 99.6 & 98 & 0 & 95.8\\
				&  CPR$_{ 89.6}$ & \textbf{100} & 98.7 & \textbf{100} & \textbf{100} & 98.6 & \textbf{100} & 100 & 100 & \textbf{100} & 92 & 96.5 & 98.9 & 95.5\\ 
				&  CPR$_{ JR-SCF}$ & \textbf{100} & 99.6 & \textbf{100} & \textbf{100} & 99 & \textbf{100} & 100 & 100 & \textbf{100} & 98.8 & 97.7 & 99.1 & 98.2 \\
				&  CPR$_{ NL4-3}$ & \textbf{100} &  100 & \textbf{100} & \textbf{100} & 98.9 & \textbf{100} & 100 & 99.8 & \textbf{100} & 100 & 96.3 & 98.8 & 100\\
				&  CPR$_{ YU2}$ & \textbf{100} & 99.7 & \textbf{100} & \textbf{100} & 99.2 & \textbf{100} & 99.5 & 99.7 & \textbf{100} & 100 & 0 & 98.6 & 99.2
				\\ \bottomrule
		\end{tabular}}
	}
\end{table*}

\subsection{Performance comparison on biallelic Solanum Tuberosum semi-experimental data}
We first evaluate performance of GAEseq on realistic simulations which, for convenience and to distinguish from perhaps more rich 
synthetic and experimental data discussed in supplementary documents, we refer to as "semi-experimental data". The 
semi-experimental data is obtained by simulating mutations, shotgun sequencing procedure, read alignment and SNP calling steps 
in a fictitious experiment on a single individual \textit{Solanum Tuberosum} (polyploid with $k=4$). Details on how exactly the 
semi-experimental data is generated and processed can be found in Supplementary Document C. We compare the performance of 
GAEseq on this data with publicly available software HapCompass (\citealp{agui12}), an algorithm that relies on graph-theoretic models to perform haplotype assembly, H-PoP (\citealp{xie16}), a dynamic programming method, and AltHap (\citealp{hash18}), a method based on tensor factorization. The performance of different methods is evaluated in terms of the MEC score and CPR. All the considered softwares were executed with their default settings, i.e. we follow instructions in the papers they were originally proposed; there are no parameter tuning steps required for these methods. We report the MEC scores and CPR achieved by the considered algorithms in Table~1. For each sequencing coverage, the mean and standard deviation (SD) of the adopted metrics are evaluated over 10 samples. As shown in the table, GAEseq achieves the lowest average MEC score as well as the lowest standard deviation of the MEC score at all sequencing coverage settings. Moreover, GAEseq achieves the highest average CPR at all coverage settings. Note that the MEC score increases with sequencing coverage since higher coverage implies more reads. The results demonstrate that the adopted graph abstraction enables GAEseq to achieve high accuracy of the reconstruction task by learning posterior probabilities 
	of the origins of reads. Fig.~3 shows the precision-recall curves for data with coverage 15$\times$, 25$\times$ and 35$\times$. Note that GAEseq performs very accuratly at high 
	sequencing coverage while its performance deteriorates at low coverage. An extended version of Table~1 with additional coverage settings is in Supplementary Document C.

We further test the performance of GAEseq on simulated biallelic diploid, polyallelic triploid and tetraploid data, and on real Solanum Tuberosum data; in addition to H-Pop, AltHap and HapCompass, comparisons on diploid data also include performance of HapCUT2 (\citealp{edge17}). GAEseq outperforms all the considered algorithms by achieving lower MEC score and higher CPR. Further details can be found in Supplementary Document D and E.

\subsection{Performance comparison on gene-wise reconstruction of real HIV-1 data}

The real HIV-1 data with pairwise distances between $2.61\%-8.45\%$ and relative frequencies between $10\%$ and $30\%$ is an \emph{in vitro} viral population of 5 known HIV-1 strains generated by Illumina's MiSeq Benchtop Sequencer (\citealp{di2014full}). These reads are then aligned to the HIV-$1_{HXB2}$ reference genome. According to (\citealp{di2014full}), we remove reads of length lower than 150bp and mapping quality scores lower than 60 for better results. We compare the performance of GAEseq on gene-wise reconstruction of the HIV population to that of other state-of-the-art methods such as PredictHaplo (\citealp{prabhakaran2014hiv}), TenSQR (\citealp{ahn2018viral}) and aBayesQR (\citealp{ahn2017abayesqr}), following their default settings. For fair benchmarking, we use the same dataset as (\citealp{ahn2018viral}) which is why the results of our benchmarking tests match those in (\citealp{ahn2018viral}). The correct phasing rate and the inferred strain frequencies are evaluated for all reconstructed strains because the ground truth for the 5 HIV-1 strains is available at (https://bmda.dmi.unibas.ch/software.html). Following (\citealp{ahn2018viral}), we evaluate predicted proportion by setting the parameter $\eta$ needed to detect the number of HIV-1 strains to $0.09$. The results in Table~2 show that GAEseq perfectly reconstructs all 5 HIV-1 strains in 8 genes while other methods correctly reconstruct components in 5 or 6 genes. This demonstrates that GAEseq's inference of read origins based on posterior probabilities enables high accuracy of the reconstruction tasks. Regarding the 5 genes where GAEseq and other methods do not achieve perfect reconstruction (p24, vpu, gp120, gp41, nef): closer examination of viral strains reconstructed by various methods suggests translocations of short viral segments within those 5 genes in the ``gold standard" dataset created by (\citealp{di2014full}). Those short translocation cause mismatch between the actual ground truth and the sequences (\citealp{di2014full}) generated. Further results on reconstruction of HIV viral communities can be found in Supplement Document F.

\section{Conclusions}

In this article, we introduce auto-encoders to the problem of reconstructing components of a genomic mixture from 
high-throughput sequencing data that is encountered in haplotype assembly and analysis of viral communities. In 
particular, a graph auto-encoder is trained to group together reads that originate from the same component of a genomic mixture and impute missing information in the SNP fragment matrix by learning from the available data. The graph convolutional encoder attempts to discover origin of the reads while the decoder aims to reconstruct haplotypes and impute missing information, effectively correcting sequencing errors. Studies on semi-experimental data show that GAEseq can achieve significantly lower MEC scores and higher CPR than the competing methods. Benchmarking tests on simulated and experimental data demonstrate that GAEseq maintains good performance even at low sequencing coverage. Studies on real HIV-1 data illustrate that GAEseq outperforms existing state-of-the-art methods in viral quasispecies reconstruction.

\small
\bibliographystyle{aaai}
\bibliography{GAEseq}

\subsection*{Supplementary Document A : Computational settings}
The models were implemented on a 3.70GHz Intel i7-8700K processor, 2 NVIDIA GeForce GTX 1080Ti computer graphics cards and 32GB RAM. Randomness of the initial weights of the auto-encoder may cause the neural network to remain in a local minimum during training. To overcome this, we run GAEseq multiple times and choose the result with the lowest MEC score. Given a SNP fragment matrix, we run GAEseq 200 times, train 200 models and get 200 reconstructed haplotype matrix candidates; the algorithm selects the candidate corresponding to the lowest MEC score automatically. In the GAEseq software, users can specify how many times to run the algorithm; the software will run automatically as opposed to manually running GAEseq multiple times.

As for hyperparameters, the number of graph convolutional layers in the auto-encoder was set to 2 (in particular, one 
read-nodes-to-SNP-nodes layer and one SNP-nodes-to-read-nodes layer); dimension of messages reduces linearly during 
message passing between the layers. For example, if the number of reads is $m$, the haplotype length is $n$,  
the ploidy is $k$ and we use 2 graph convolutional layers and a dense layer, the dimensions of the weight and bias matrix 
of the first and second layer are $n \times \left \lceil{(n - \frac{n - k}{3})}\right \rceil$ and 
$\left \lceil{(n - \frac{n - k}{3})}\right \rceil \times \left \lceil{(n - 2 \times \frac{n - k}{3})}\right \rceil$, respectively. The 
dimensions of the weight and bias matrix of the denser layer are set to 
$\left \lceil{(n - 2 \times \frac{n - k}{3})}\right \rceil \times k$ and $m \times k$, respectively. We use the Adam optimizer 
(\citealp{adam2015}), set the step size to 0.0001, and set all other parameters to their default values in Tensorflow. We also use Xavier initialization (\citealp{Xavier2010}) with default 
settings and the number of epoches set to 100. In our studies, we found that an architecture with two graph convolutional 
layers achieves significantly better results than state-of-the-art haplotype assembly methods. To ensure the model 
generalizes well to unobserved entries, we added dropout regularization to message passing between the layers; for 
each layer, the dropout probability is 0.1. For all experiments on viral quasispecies reconstruction, the initial number of 
clusters $k_0$ is set to 2.

\subsection*{Supplementary Document B : Determining the number of components in a viral quasispecies}

When reconstructing haplotypes sampled by a collection of sequencing reads, GAEseq requires as input the number of haplotypes, 
$k$. While the ploidy of an individual organism in the haplotype assembly problem is known a priori, cardinality of a viral community 
needs to be estimated. To determine $k$, we examine the improvement rate of the MEC score defined as
\begin{equation}
\mbox{MECimpr}(k) = \frac{\text{MEC}(k)-\text{MEC}(k+1)}{\text{MEC}(k)}.
\end{equation}
Recall that the MEC score is defined as the smallest number of the observed entries in $R$ that need to be altered such that the resulting 
data is consistent with having originated from $k$ distinct haplotypes. The score decreases monotonically with $k$; however, 
once $k$ reaches the actual number of components, the improvement rate of the MEC score ($\mbox{MECimpr}$) saturates. 
To find the saturation point, we compare $\mbox{MECimpr}$ with a pre-defined threshold. Following 
(\citealp{ahn2018viral}), the number of components is determined via binary search. Specifically, starting from an initial $k_0$, the number 
of components is updated as $k_{\tau} \leftarrow 2k_{\tau-1}$ until $\mbox{MECimpr}(k_\tau) \le \eta$; at this point, the number of 
components starts to decrease as $k_{\tau+1} \leftarrow \lfloor(k_{\tau} + \max \{1,k_i\} )/ 2 \rfloor$ where $\{ i\in \{1,\cdots,\tau-1\}: 
k_i \le k_\tau \}$. Once $\mbox{MECimpr}(k_\tau)>\eta$, the number of components increases again as 
$k_{\tau+1} \leftarrow \lfloor(k_{\tau} + \min k_i )/2\rfloor$ where $\{ i \in \{1,\cdots,\tau-1\} : k_i > k_\tau \}$.
If $k_{\tau} = k_{\tau-1}$, the search procedure stops by assigning $k_{\tau+1} \leftarrow k_{\tau}+1$ 
which is the estimated number of strains. The recommended choice of the threshold $\eta$
is discussed in (\citealp{ahn2017abayesqr}) where the estimation of the number of components via $\mbox{MECimpr}$ was demonstrated to be robust with respect to the choice of the threshold.

\subsection*{Supplementary Document C : Performance comparison on biallelic Solanum Tuberosum semi-experimental data}

The semi-experimental data is obtained by simulating mutations, shotgun sequencing procedure, read alignment and SNP calling 
steps in an experiment on a single individual \textit{Solanum Tuberosum}. In particular, we use \textit{Haplogenerator} (\citealp{motazedi2018exploit}) to generate haplotypes by introducing independent mutations that follow the lognormal distribution of a randomly selected genome region from \textit{Solanum Tuberosum} chromosome 5 (\citealp{potato2011genome}) of length 5000 bp. The mean distance between neighboring SNPs and the standard deviation (SD) are set to 21 bp and 27 bp, respectively, as previously suggested by (\citealp{motazedi2018exploit}). Due to \textit{Haplogenerator}'s limitations, we constrain mutations to transitions and do not consider transversions (i.e., mutations are constrained to be between A and C and between G and T). $2 \times 250$ bp-long Illumina's MiSeq reads of inner distance 50 bp and standard deviation 10 bp are generated to uniformly sample haplotypes using \textit{ART} software (\citealp{huang2012art}) with default setting. Following this step, the generated reads are aligned to the reference genome using the BWA-MEM algorithm (\citealp{li2009fast}); the reads having mapping quality score lower than 60 or being shorter than 70 bp are discarded. SNPs are called if, at any given site, the abundance of a minor allele exceeds a predetermined threshold; the SNP fragment matrix is formed by collecting all such heterozygous sites. Seven different sets of semi-experimental data obtained by sampling at varying coverage (10$\times$, 15$\times$, 20$\times$, 25$\times$, 30$\times$, 35$\times$ and 40$\times$) are generated; each set consists of 10 samples. We first generate genome regions of length 5000 bp by partitioning the \textit{Solanum Tuberosum} chromosome 5 and then randomly select 70 among them (generated haplotypes and reads are different for each sample). The sequencing error rate is automatically set by the built-in quality profiles of \textit{ART} inferred from large amounts of recalibrated sequencing data (\citealp{huang2012art}). Table~1 shows the performance comparison of GAEseq, AltHap, HapCompass and H-PoP on biallelic Solanum Tuberosum semi-experimental data. 

\begin{table*}[h!]
	\caption{Performance comparison of GAEseq, AltHap, HapCompass and H-PoP on biallelic Solanum Tuberosum 
		semi-experimental data.}
	\centering\label{tab:01}
	{\begin{tabular}{ p{0.13\textwidth} p{0.13\textwidth} p{0.13\textwidth} p{0.13\textwidth} p{0.13\textwidth} p{0.13\textwidth}}\toprule 
			& & \multicolumn{2}{l}{MEC} & \multicolumn{2}{l}{CPR}\\
			Coverage & & Mean & SD & Mean & SD\\
			\hline 
			\multirow{4}{*}{10}& GAEseq & \textbf{18.500} & \textbf{4.552} & \textbf{0.848} & 0.074\\
			& HapCompass & 100.300 &43.584 & 0.769 & \textbf{0.039}\\
			& H-PoP & 19.700 & 25.254 & 0.803 & 0.086\\
			& AltHap & 64.100 & 32.953 & 0.727 & 0.072\\
			\hline
			\multirow{4}{*}{15}& GAEseq & \textbf{8.200} & \textbf{4.686} & \textbf{0.822} & 0.048\\
			& HapCompass & 100.700 & 66.150 & 0.763 & \textbf{0.046}\\
			& H-PoP & 28.700 & 32.667 & 0.783 & 0.066\\
			& AltHap & 59.100 & 28.125 & 0.709 & 0.054\\
			\hline
			\multirow{4}{*}{20}& GAEseq & \textbf{16.800} & \textbf{15.873} & \textbf{0.862} & 0.062\\
			& HapCompass & 95.600 & 53.883 & 0.795 & \textbf{0.047}\\
			& H-PoP & 30.500 & 37.023 & 0.791 & 0.078\\
			& AltHap & 82.100 & 56.658 & 0.737 & 0.068\\
			\hline
			\multirow{4}{*}{25}& GAEseq & \textbf{8.400} & \textbf{4.719} & \textbf{0.831} & 0.081\\
			& HapCompass & 124.800 & 132.156 & 0.810 & 0.063\\
			& H-PoP & 33.800 & 47.434 & 0.798 & \textbf{0.046}\\
			& AltHap & 92.600 & 83.649 & 0.756 & 0.068\\
			\hline
			\multirow{4}{*}{30}& GAEseq & \textbf{27.200} & \textbf{19.887} & \textbf{0.914} & \textbf{0.033}\\
			& HapCompass & 306.800 & 187.934 & 0.796 & 0.081\\
			& H-PoP & 34.200 & 32.798 & 0.879 & 0.088\\
			& AltHap & 263.000 &499.659 & 0.762 & 0.133\\
			\hline
			\multirow{4}{*}{35}& GAEseq & \textbf{10.700} & \textbf{3.234} & \textbf{0.857} & 0.087\\
			& HapCompass & 217.400 & 174.135 & 0.775 & \textbf{0.072}\\
			& H-PoP & 41.700 & 53.971 & 0.823 & 0.094\\
			& AltHap & 164.000 & 101.583 & 0.754 & 0.093\\
			\hline
			\multirow{4}{*}{40}& GAEseq & \textbf{16.400} & \textbf{7.333} & \textbf{0.835} & \textbf{0.034}\\
			& HapCompass & 208.000 & 176.699 & 0.833 & 0.070\\
			& H-PoP & 30.4 & 28.487 & 0.823 & 0.102\\
			& AltHap & 195.8 & 281.641 & 0.762 & 0.084\\
			\bottomrule
	\end{tabular}}
\end{table*}

\subsection*{Supplementary Document D : Performance comparison on simulated biallelic diploid data and polyallelic triploid and tetraploid data.}
To further test GAEseq, we evaluate its performance on synthetic data. Once again we use \textit{Haplogenerator} (\citealp{motazedi2018exploit}) to generate haplotypes of a randomly synthesized reference genome of length 5000 bp. The mean distance between neighboring SNPs and the standard deviation (SD) are set to 5 bp and 3 bp respectively, creating haplotype blocks of length about 500. All the possible mutations were allowed and set to be equally likely, leading to not only biallelic but also polyallelic SNPs in the synthesized haplotype data. Illumina's MiSeq read generation, read alignment and SNP calling procedures are implemented following the same procedure as in the case of semi-experimental data from Section~3.1. The data synthesized in this fashion consists of $24$ different sets, each with 10 samples, as we explore different ploidy (k = 2, 3 and 4) and sequencing coverage (5$\times$, 10$\times$, 15$\times$, 20$\times$, 25$\times$, 30$\times$, 35$\times$ and 40$\times$).

\begin{table*}[h!]
\caption{Performance comparison of GAEseq, HapCut2, HapCompass,  H-PoP and AltHap on simulated biallelic diploid data.}
\centering\label{tab:02}
{\begin{tabular}{ p{0.13\textwidth} p{0.13\textwidth} p{0.13\textwidth} p{0.13\textwidth} p{0.13\textwidth} p{0.13\textwidth}}\toprule 
 & & \multicolumn{2}{l}{MEC} & \multicolumn{2}{l}{CPR}\\
Coverage & & Mean & SD & Mean & SD\\
\hline 
 \multirow{4}{*}{5}& GAEseq & \textbf{23.300} & \textbf{4.165}  & \textbf{0.996}  & \textbf{0.002} \\
 & HapCUT2 & 110.500  & 23.922 &0.975 & 0.006 \\
 &HapCompass	 &87.500 &	25.903	 &0.965	 &0.010 \\
 &H-Pop &	40.000	 &30.551	 &0.989 &	0.011 \\
 & AltHap & 884.200 & 659.565 & 0.699 & 0.204\\
\hline 
 \multirow{4}{*}{10}& GAEseq & \textbf{30.700} & \textbf{6.667} & \textbf{0.999} & \textbf{0.001}\\
 & HapCUT2 &213.600& 63.132 &  0.980 &  0.005\\
 &HapCompass	 &159.600 &	58.329 &	0.974 &	0.005\\
 &H-Pop	 &34.600	 &6.736 &	0.997	 &0.004\\
 & AltHap & 583.900 & 948.344 & 0.796 & 0.218\\
\hline
 \multirow{4}{*}{15}& GAEseq & \textbf{47.800} & \textbf{8.587} &\textbf{0.999} & \textbf{0.001}\\
 & HapCUT2 & 339.800  &  59.066 & 0.978  & 0.003 \\
 &HapCompass &	268.300 &	67.003	 &0.971 & 0.005\\
 &H-Pop	 &47.900	 &9.539 &	0.998 &0.002\\
 & AltHap & 342.900 & 379.213 & 0.852 & 0.169\\
 \hline
 \multirow{4}{*}{20}& GAEseq & \textbf{70.900} & \textbf{10.754} & \textbf{1.000} & \textbf{0.001}\\
 & HapCUT2 & 519.400  & 57.386  & 0.972  & 0.010 \\
 &HapCompass	 &408.000 &	81.067 &	0.961	 &0.018 \\
 &H-Pop	 &129.700 &	191.788 &	0.989 &	0.030 \\
 & AltHap & 668.400 & 579.261 & 0.787 & 0.201\\
 \hline
 \multirow{4}{*}{25}& GAEseq & \textbf{85.200} & \textbf{16.130} & \textbf{1.000} & \textbf{0.001}\\
 & HapCUT2 &  613.000 &  157.786 & 0.977  & 0.006 \\
  &HapCompass	&460.700 &	97.637 &	0.968	 &0.007 \\
  &H-Pop	 &85.700	 &17.192	 &0.998 &	0.003 \\
 & AltHap & 1151.600 & 649.058 & 0.743 & 0.150\\
 \hline
 \multirow{4}{*}{30}& GAEseq & \textbf{97.800} & 8.954 & \textbf{1.000} & \textbf{0.000}\\
 & HapCUT2 & 685.300  &  180.714  &  0.979 &  0.006 \\
  &HapCompass	&591.600&	150.400	&0.968	&0.009\\
  &H-Pop	 &98.000&	\textbf{8.743}&	0.999&	0.001\\
 & AltHap & 554.000 &612.292 & 0.871 & 0.185\\
 \hline
 \multirow{4}{*}{35}& GAEseq & \textbf{107.300} & 8.138 & \textbf{1.000} & \textbf{0.001}\\
 & HapCUT2 &  827.600 & 202.643  & 0.978  &  0.006\\
  &H-Pop	 &702.200	&180.647	&0.968&	0.007\\
  &H-Pop	 &107.900&	\textbf{8.006}&	0.999	&0.001\\
 & AltHap & 668.800 & 730.814 & 0.891 & 0.146\\
 \hline
 \multirow{4}{*}{40}& GAEseq & \textbf{124.000} & 10.499 & \textbf{1.000} & \textbf{0.001}\\
 & HapCUT2 & 1015.400  &  219.442 & 0.977  & 0.006 \\
&HapCompass	&896.500	&204.603	&0.965&	0.008 \\
&H-Pop	&124.500	&\textbf{10.277}	&0.999	&\textbf{0.001} \\
 & AltHap & 1073.300 & 1099.181 & 0.847 & 0.184\\
\bottomrule
\end{tabular}}
\end{table*}

\begin{table*}[hbt!]
\caption{Performance comparison of GAEseq and AltHap on simulated polyallelic triploid data.}
\centering\label{tab:03}
{\begin{tabular}{ p{0.13\textwidth} p{0.13\textwidth} p{0.13\textwidth} p{0.13\textwidth} p{0.13\textwidth} p{0.13\textwidth}}\toprule 
 & & \multicolumn{2}{l}{MEC} & \multicolumn{2}{l}{CPR}\\
Coverage & & Mean & SD & Mean & SD\\
\hline 
 \multirow{3}{*}{5}& GAEseq & \textbf{103.400} & \textbf{51.379} & \textbf{0.920} & \textbf{0.047}\\
 & AltHap & 1908.500 & 237.324 &0.559 & 0.059\\
\hline 
 \multirow{2}{*}{10}& GAEseq & \textbf{112.800} & \textbf{45.917} & \textbf{0.958} & \textbf{0.037}\\
 & AltHap & 1769.300 & 948.754 &0.760 & 0.091\\
\hline
 \multirow{2}{*}{15}& GAEseq & \textbf{165.800} & \textbf{106.999} & \textbf{0.945} & \textbf{0.073}\\
 & AltHap & 1058.100 & 864.563 & 0.796 & 0.123\\
 \hline
 \multirow{2}{*}{20}& GAEseq & \textbf{241.300} & \textbf{159.657} & \textbf{0.959} & \textbf{0.047}\\
 & AltHap & 1287.100 & 578.507 & 0.682 & 0.070\\
 \hline
 \multirow{2}{*}{25}& GAEseq & \textbf{314.900} & \textbf{158.326} & \textbf{0.934} & \textbf{0.070}\\
 & AltHap & 1430.200 & 757.482 & 0.775 & 0.093\\
 \hline
 \multirow{2}{*}{30}& GAEseq & \textbf{292.400} & \textbf{203.242} & \textbf{0.974} & \textbf{0.040}\\
 & AltHap & 2133.200 &1082.576 & 0.729 & 0.109\\
 \hline
 \multirow{2}{*}{35}& GAEseq & \textbf{306.200} & \textbf{196.918} & \textbf{0.982} & \textbf{0.037}\\
 & AltHap & 2928.700 & 869.617 & 0.723 & 0.075\\
 \hline
 \multirow{2}{*}{40}& GAEseq & \textbf{502.200} & \textbf{247.380} & \textbf{0.922} & \textbf{0.088}\\
 & AltHap & 2943.600 & 1113.480 & 0.737 & 0.104\\
\bottomrule
\end{tabular}}
\end{table*}

\begin{table*}[hbt!]
\caption{Performance comparison of GAEseq and AltHap on simulated polyallelic tetraploid data.}
\centering\label{tab:04}
{\begin{tabular}{ p{0.13\textwidth} p{0.13\textwidth} p{0.13\textwidth} p{0.13\textwidth} p{0.13\textwidth} p{0.13\textwidth}}\toprule 
 & & \multicolumn{2}{l}{MEC} & \multicolumn{2}{l}{CPR}\\
Coverage & & Mean & SD & Mean & SD\\
\hline 
 \multirow{3}{*}{5}& GAEseq & \textbf{266.700} & \textbf{46.371} & \textbf{0.739} & \textbf{0.041}\\
 & AltHap & 2641.700 & 410.159 & 0.544 & 0.056 \\
\hline 
 \multirow{2}{*}{10}& GAEseq & \textbf{415.100} & \textbf{74.608} & \textbf{0.800} & \textbf{0.051}\\
 & AltHap & 2807.200 & 938.668 & 0.658 & 0.075\\
\hline
 \multirow{2}{*}{15}& GAEseq & \textbf{592.200} & \textbf{112.282} & \textbf{0.798} & \textbf{0.054}\\
 & AltHap & 2742.500 & 1055.672 & 0.718 & 0.081\\
 \hline
 \multirow{2}{*}{20}& GAEseq & \textbf{628.900} & \textbf{245.841} & \textbf{0.843} & \textbf{0.047}\\
 & AltHap & 1929.700 & 1008.766 & 0.729 & 0.063\\
 \hline
 \multirow{2}{*}{25}& GAEseq & \textbf{881.900} & \textbf{189.987} & \textbf{0.845} & \textbf{0.058}\\
 & AltHap & 1987.100 & 1091.893 & 0.779 & 0.084\\
 \hline
 \multirow{2}{*}{30}& GAEseq & \textbf{944.100} & \textbf{182.440} & \textbf{0.848} & \textbf{0.041}\\
 & AltHap & 2265.200 & 1277.366 & 0.759 & 0.051\\
 \hline
 \multirow{2}{*}{35}& GAEseq & \textbf{815.900} & \textbf{295.195} & \textbf{0.866} & \textbf{0.063}\\
 & AltHap & 3906.400 & 1131.654 & 0.747 & 0.056\\
 \hline
 \multirow{2}{*}{40}& GAEseq & \textbf{949.500} & \textbf{319.238} & \textbf{0.878} & \textbf{0.046}\\
 & AltHap & 3775.300 & 1036.702 & 0.762 & 0.075\\
\bottomrule
\end{tabular}}
\end{table*}

For the diploid synthetic data sets, we represent an allele by 0 if it coincides with the corresponding reference allele and by 
1 if it is an alternative allele. SNP positions with only alternative alleles are removed.  In addition to H-PoP, HapCompass and AltHap, we also compare GAEseq with HapCUT2 (\citealp{edge17}); by design, use of HapCUT2 is limited to haplotype assembly of diploids. The metrics of performance are the previously introduced MEC score and CPR. Table 2 shows the mean and standard deviation of the MEC score and CPR for diploid data. The results are evaluated over 10 samples for each combination of ploidy and coverage.  GAEseq achieves the lowest average 
MEC score and the lowest standard deviation of the MEC score for almost all coverage settings; its performance is followed by those of H-PoP, HapCompass, HapCut2 and AltHap. The average CPR achieved by GAEseq is very close to 1 for all coverage settings, indicating that GAEseq is able to near-perfectly reconstruct haplotypes of diploid species even when the coverage is very low; its performance is followed by those of H-PoP, HapCut2, HapCompass and AltHap. When the coverage is $20\times$, the average CPR achieved by GAEseq is $100\%$ while it is approximately $98.9\%$, $97.2\%$, $96.1\%$ and $74.3\%$ for H-PoP, HapCut2, HapCompass and AltHap, respectively.

For the polyploid synthetic data sets, both H-PoP and HapCompass are restricted to reconstruction of biallelic haplotypes and are not applicable to the assembly of polyallelic ones. Furthermore, recall that HapCUT2 can only be applied to diploid haplotypes. We therefore limit performance comparison of GAEseq on polyploid synthetic data to only AltHap; Tables 3 and 4 illustrate the mean and standard deviation of the MEC score and CPR for triploid and tetraploid data, respectively. The results are evaluated over 10 samples for each combination of ploidy and coverage. As can be seen in these tables, GAEseq outperforms AltHap for all ploidy and coverage settings.  As shown in Table 3, GAEseq performs well on triploid data, achieving $92\%$ average CPR and relatively small standard deviation even for the low coverage of $5\times$; at the same time, performance of AltHap deteriorates rapidly with increased ploidy, achieving $72\%$ average CPR while GAEseq achieves $98.2\%$ at coverage $30\times$. As illustrated in Table 4, in applications to tetraploid data the performance of GAEseq starts to gracefully deteriorate -- when the coverage is $10\times$, GAEseq achieves average CPR of approximately $80\%$ while in the same scenario AltHap achieves average CPR of approximately $65\%$. When the coverage is increased to $40\times$,  GAEseq achieves average CPR of approximately 
$87.8\%$ while AltHap achieves average CPR of approximately $76.2\%$.

\subsection*{Supplementary Document E : Performance comparison on real Solanum Tuberosum data}
We further test the performance of GAEseq on real potato data (accession SRR6173308) at \textit{Solanum Tuberosum} chromosome 5 (\citealp{potato2011genome}). The 10 samples of real potato data are generated by first randomly selecting 10 genome regions of length varying from 5032 to 7573 and then aligning the Illumina HiSeq 2000 paired-end reads to the selected genome regions. After the read alignment step using the BWA-MEM algorithm (\citealp{li2009fast}), the SNP calling step is implemented to create the SNP fragment matrix. Reads having mapping quality score lower than 60 or shorter than 70 bp are discarded. Since the ground truth haplotypes are not available for this dataset, we only evaluate the performance of GAEseq and the competing methods in terms of the MEC score. Table 5 compares the performance of GAEseq, AltHap, HapCompass and H-PoP averaged over 10 selected regions of the real Solanum Tuberosum data.
As can be seen from the table, GAEseq outperforms all the competing schemes in terms of both the average MEC score and its standard deviation, achieving 379.8 average MEC score. GAEseq is followed by H-PoP and AltHap while HapCompass achieves the highest average MEC score.

\begin{table}[h]
\caption{Performance comparison of GAEseq, AltHap, HapCompss and H-PoP on the real Solanum Tuberosum data.}
\centering\label{tab:05}
{\begin{tabular}{p{0.1\textwidth} p{0.1\textwidth} p{0.1\textwidth}}\toprule 
 & \multicolumn{2}{l}{MEC}\\
 & Mean & SD \\
 \hline
 GAEseq & \textbf{379.8} & \textbf{271.61} \\
HapCompass & 2726 &	2393.7\\
H-PoP & 409.5 & 282.24\\
 AltHap & 742.1 & 469.5
\\ \bottomrule
\end{tabular}}
\end{table}

\subsection*{Supplementary Document F : Further results on reconstruction of HIV viral communities}
\label{HIV_gagpol}

Table 6 shows the gene-wise reconstruction results on the real HIV-1 data that include inferred frequencies
(omitted from Table~2 in the main paper for brevity).

We further evaluate the performance of GAEseq on the 4036bp long gag-pol region. Following (\citealp{ahn2018viral}), we divide the gag-pol region into overlapping blocks, reconstruct the viral components in each block independently, and combine the results to reconstruct the full region of interest. Specifically, the region is divided into a sequence of blocks of length 500bp where the consecutive blocks overlap by 250bp. We run GAEseq to perform reconstruction of viral components in each of the total 18 blocks and merge the results to retrieve the entire region of interest. Particularly, the mismatches between strains reconstructed on two consecutive blocks in the overlapping region are corrected based on majority voting using reads that are covering the mismatched positions and are assigned to the aligned strains. Following this procedure, GAEseq perfectly reconstructed all of 5 HIV-1 strains in the gag-pol region, achieving 100\% Reconstruction Rate for all 5 strains and Predicted Proportion of 1 on 355241 remained paired-end reads. The frequencies of 5 HIV-1 strains are estimated as $15.21\%, 19.34\%, 25.56\%, 27.61\%$ and $12.27\%$ by counting the proportion of reads assigned to the same strain; these results are consistent with the frequencies estimated by aBayesQR and TenSQR softwares. 

\begin{sidewaystable*}[h!]
	\centering
	\bigskip
	\centering\small\setlength\tabcolsep{2pt}
	\caption{Performance comparisons of GAEseq, TenSQR, aBayesQR and PredictHap on a real HIV-1 5-virus-mix data.}
	\resizebox{1.01 \textwidth}{!}{
		\begin{tabular}{c c c c c c c c c c c c c c c }\toprule 
			& &  p17 &  p24&  p2-p6 &  PR&  RT &  RNase & int &  vif &  vpr &  vpu &  gp120 &  gp41 &  nef\\
			\hline
			GAEseq &  PredProp & \textbf{1} & 1 & \textbf{1} & \textbf{1} & \textbf{1.2} & \textbf{1} & \textbf{1} & \textbf{1} & \textbf{1} & 1.2 & 1 & 1 & 1\\
			&  CPR$_{ HXB2}$ &  \textbf{100(20.5)} & 99.4(17.1) & \textbf{100(21)} & \textbf{100(30.9)} & \textbf{100(12.1)} & \textbf{100(9.6)} & \textbf{100(13.6)}  & \textbf{100(10.4)}  & \textbf{100(6.6)} & 100(34.3) & 96.2(8.7) & 96.7(2.8)  & 100(6.6) \\
			&  CPR$_{ 89.6}$& \textbf{100(18.8)} & 99.4(21.8) & \textbf{100(20)} & \textbf{100(18)} & \textbf{100(18.2)} & \textbf{100(20.9)} & \textbf{100(18.2)} & \textbf{100(20.4)}  & \textbf{100(20.3)} & 99.2(10.5) & 99.4(24.1) & 100(25.6)  & 98.2(22.9)\\
			&  CPR$_{ JR-SCF}$ & \textbf{100(30.9)} & 100(31.5) & \textbf{100(27)} & \textbf{100(21.6)} & \textbf{100(23.5)} & \textbf{100(20.2)} & \textbf{100(21.5)}  & \textbf{100(29.8)}  & \textbf{100(34.6)} & 100(36.4) & 99.9(33.0) & 100(27)  & 99.3(19.7)\\
			&  CPR$_{ NL4-3}$ & \textbf{100(17.4)} & 100(18.3) & \textbf{100(14.1)} & \textbf{100(19.7)} & \textbf{100(30.6)} & \textbf{100(33.1)} & \textbf{100(37.1)}  & \textbf{100(32.8)}  & \textbf{100(30.6)} & 100(7.9) & 100(29.6) & 100(34.5)  & 99.8(39.1)\\
			&  CPR$_{ YU2}$ & \textbf{100(12.3)} &  100(11.2) & \textbf{100(18)} & \textbf{100(9.9)} & \textbf{100(11.9)} & \textbf{100(16.2)} & \textbf{100(9.6)}  & \textbf{100(6.6)}  & \textbf{100(7.9)} & 100(10.2) & 99.6(4.6) & 100(10)  & 98.1(11.7)\\
			PredictHap &  PredProp &  \textbf{1} &  0.6 &  \textbf{1} &  \textbf{1} &  \textbf{1} &  0.8 &  0.8 & 0.8 & \textbf{1} & 0.8 & 0.8 & 0.8 & 0.8\\
			&  CPR$_{ HXB2}$ & \textbf{100(17.8)}  & 0(0)  & \textbf{100(18.7)} & \textbf{100(15.2)} & \textbf{100(12.2)} & 98.9(25.4)  & 100(12.1) & 100(17.7) & \textbf{100(10.2)} & 93.2(10.8) & 0(0) & 0(0) & 0(0) \\
			&  CPR$_{ 89.6}$ & \textbf{100(19.9)} & 100(46.4) & \textbf{100(21.7)} & \textbf{100(22.2)} & \textbf{100(19.4)} & 100(18.2) & 99.8(27.6) & 100(20.9) & \textbf{100(22.1)} & 0(0) & 97.8(20.7) & 100(26.7) & 98.8(20.7)\\
			&  CPR$_{ JR-SCF}$ & \textbf{100(31.9)} &  100(21.8) & \textbf{100(30.3)} & \textbf{100(26.9)} & \textbf{100(23.4)} & 100(23.2) & 100(22.3) & 100(24.9) & \textbf{100(23.7)} & 100(34.1) & 99.7(42.7) & 100(28.9) & 100(23.2)\\
			&  CPR$_{ NL4-3}$ & \textbf{100(17)} & 99.1(31.8) & \textbf{100(16.4)} & \textbf{100(20.9)} & \textbf{100(30.2)} & 100(33.2) & 100(38.1) & 100(36.6) & \textbf{100(35.5)} & 100(47.1) & 100(28.6) & 100(32.7) & 100(39.3)\\
			&  CPR$_{ YU2}$ & \textbf{100(13.4)} & 0(0) & \textbf{100(12.9)} & \textbf{100(14.8)} & \textbf{100(14.7)} & 0(0) & 0(0) & 0(0) & \textbf{100(8.5)} & 100(7.9) & 98.6(7.9) & 100(11.7) & 100(16.9)\\
			TenSQR &  PredProp & \textbf{1} & 1.6 & \textbf{1} & 1 & 1.4 &  \textbf{1}  &  \textbf{1}  &  \textbf{1}   &  \textbf{1}  & 1.6 & 2.2 & 1.2 & 0.8\\
			&  CPR$_{ HXB2}$ & \textbf{100(18.7)} & 98.9(13.1) & \textbf{100()17.4} & 100(9.9) & 99.2(12.1) & \textbf{100(9.2)}  & \textbf{100(8.1)}  & \textbf{100(9.6)}  & \textbf{100(7.2)}  & 92.8(5.9) & 96.0(18) & 99.0(11.5) & 0(0)\\
			&  CPR$_{ 89.6}$ & \textbf{100(18.4)} & 100(19.6) & \textbf{100(20.1)} & 100(17.2)  & 98.0(13.5) & \textbf{100(17.2)}  & \textbf{100(16.7)}  & \textbf{100(25)}  & \textbf{100(19.3)}  & 94.0(15) & 97.2(10.3) & 100(27.8) & 95.7(26)\\
			&  CPR$_{ JR-SCF}$ & \textbf{100(33.8)} & 100(33) & \textbf{100(33.6)} & 100(21.7) & 100(20.7) & \textbf{100(24.6)} & \textbf{100(23.3)} & \textbf{100(20.5)} & \textbf{100(20.3)} & 100(31.4) & 98.3(33.5) & 97.7(18.8) & 99.8(19)\\
			&  CPR$_{ NL4-3}$ &  \textbf{100(17)}& 99.3(19.7) & \textbf{100(17.2)} & 100(21.4) & 99.5(26.7) & \textbf{100(37.7)} & \textbf{100(41.2)} & \textbf{100(38.4)} & \textbf{100(46.2)} & 100(38.8) & 99.8(9.2) & 99.5(23.2) & 99.7(42.7)  \\
			&  CPR$_{ YU2}$ & \textbf{100(12.1)} &  99.3(14.6) & \textbf{100(7.7)} & 99.7(29.8) & 99.7(14.5) & \textbf{100(11.4)} & \textbf{100(10.7)} & \textbf{100(6.5)} & \textbf{100(7.1)} & 100(4.1) & 94.9(10.5) & 100(10.2) & 98.6(12.3)\\
			aBayesQR & PredProp & \text{1} & 1 & \textbf{1}  & \textbf{1}  & 1 & \textbf{1}  & 1 & 1 & \textbf{1.2} & 1 & 0.8 & 0.8 & 1.2\\
			&  CPR$_{ HXB2}$ & \textbf{100(16.3)} & 99.4(21.1) & \textbf{100(22.2)} & \textbf{100(12.5)} & 98.5(24.3) & \textbf{100(16.1)} & 99.9(9.7) & 100(9.2) & \textbf{100(16.4)} & 99.6(17) & 98(30.3) & 0(0) & 95.8(11.4)\\
			&  CPR$_{ 89.6}$ & \textbf{100(27.1)} & 98.7(17) & \textbf{100(17.3)} & \textbf{100(17.3)} & 98.6(18.1) & \textbf{100(19.7)} & 100(22.2) & 100(20.6) & \textbf{100(16.3)} & 92(10.4) & 96.5(20.2) & 98.9(23.7) & 95.5(16.4)\\ 
			&  CPR$_{ JR-SCF}$ & \textbf{100(31.3)} & 99.6(24.6) & \textbf{100(25.8)} & \textbf{100(29.9)} & 99(21.5) & \textbf{100(22.1)} & 100(20.8) & 100(32.7) & \textbf{100(27)} & 98.8(26.7) & 97.7(21.4) & 99.1(29.7) & 98.2(21.1) \\
			&  CPR$_{ NL4-3}$ & \textbf{100(12.9)} &  100(21.6) & \textbf{100(25.6)} & \textbf{100(20.1)} & 98.9(17.7) & \textbf{100(30)} & 100(39.5) & 99.8(28.5) & \textbf{100(23.2)} & 100(41.3) & 96.3(28) & 98.8(36.6) & 100(31.8)\\
			&  CPR$_{ YU2}$ & \textbf{100(12.4)} & 99.7(15.8) & \textbf{100(9.2)} & \textbf{100(20.3)} & 99.2(18.5) & \textbf{100(12.2)} & 99.5(7.9) & 99.7(9) & \textbf{100(17.1)} & 100(4.6) & 0(0) & 98.6(10.1) & 99.2(14)
			\\ \bottomrule
		\end{tabular}
	}
	\scriptsize
	
	Predicted Proportion (PredProp) and Correct Phasing Rate (CPR ($\%$)) for GAEseq, PredictHaplo, TenSQR and aBayesQR applied to reconstruction of HIV-1HXB2, HIV-189.6,
	HIV-1JR-CSF, HIV-1NL4-3 and HIV-1YU2 for all 13 genes of the HIV-1 dataset. Frequencies are reported in parenthesis.
	
\end{sidewaystable*}

\end{document}